\documentclass[10pt, journal,twocolumn]{IEEEtran}
 \pdfoutput=1
 
\usepackage[font=small]{caption}
\usepackage{color}
\usepackage{amsmath}
\usepackage{mathtools}
\usepackage{amsmath}
\usepackage{amssymb}
\usepackage{cite}
\usepackage{array}
\usepackage{amsthm}
\usepackage{amsmath}
\usepackage{tabulary}
\usepackage{multirow}
\usepackage{subfig}
\usepackage{enumerate}
\IEEEoverridecommandlockouts
\usepackage{pifont}
\usepackage{epsfig}
\usepackage{graphicx}
\usepackage{url}
\usepackage{amssymb}
\usepackage{paralist} %
\usepackage{algorithm}
\usepackage{algpseudocode}
\usepackage{pifont}

\graphicspath{ {./figs/} }

\newcommand{\res}{\mathrm{W}}

\newcommand{\latency}{l}
\newcommand{\rate}{r}
\newcommand{\se}{c}
\newcommand{\Pse}{p}

\newcommand{\lm}{\nu}

\newcommand{\uth}[1]{#1^\text{th}}

\newcommand{\SINR}{\mathtt{SINR}}

\newcommand{\SNR}{\mathtt{SNR}}

\newcommand{\af}{\eta} 
 
\theoremstyle{definition}

\theoremstyle{thm}

\begin{document}
\title{ Proportional Fair Traffic Splitting and Aggregation  in    Heterogeneous Wireless Networks}

\author{\IEEEauthorblockN{Sarabjot Singh, Mikhail Geraseminko, Shu-ping Yeh, Nageen Himayat, Shilpa Talwar}
\thanks{The authors are with  Intel Corp., Santa Clara, USA (email: \{sarabjot.singh\}, \{shu-ping.yeh\}, \{nageen.himayat\},\{shilpa.talwar\}@intel.com). The work by the second author was done during an internship with Intel Corp.  
}}
\maketitle

\begin{abstract}
Traffic load balancing and resource allocation is set to play a crucial role in leveraging the dense and increasingly heterogeneous deployment of multi-radio wireless networks. Traffic aggregation across different access points (APs)/radio access technologies (RATs) has become an important feature of recently introduced cellular standards on LTE dual connectivity and LTE-WLAN aggregation (LWA). Low complexity traffic splitting solutions  for scenarios  where the APs are not necessarily collocated are of great interest for operators. In this paper, we consider a scenario, where traffic for each user may be split across macrocell and an LTE or WiFi small cells connected by non-ideal backhaul links, and develop a closed form solution for optimal aggregation accounting for the backhaul delay. The optimal solution lends itself to a ``water-filling" based interpretation, where the fraction of user's traffic sent over macrocell is proportional to ratio of user's peak capacity on that macrocell and its throughput on the small cell.   Using comprehensive system level simulations, the developed optimal solution is shown to provide substantial edge and median throughput gain over algorithms   representative of current 3GPP-WLAN interworking solutions. The achievable performance benefits hold promise for operators expecting to introduce aggregation solutions with their existing WLAN deployments.
\end{abstract}

\section{Introduction}
The burgeoning demand for wireless data has led to denser and heterogeneous deployment of wireless networks.  This  heterogeneity manifests itself in terms of networks differing in coverage area per AP, RATs, propagation characteristics,  backhaul delay, etc.  As a result, more often than not,  user equipment (UE) would lie in the overlapping coverage areas of multiple APs/RATs. Techniques to optimally leverage such simultaneous availability of multiple RATs alongwith the multi-link traffic aggregation capability of user equipments (UEs) are set to play crucial role in the next generation of wireless networks \cite{And5G14}. 

Load balancing in multi-RAT heterogeneous networks (HetNets) via intelligent  UE association has attracted significant interest from both academia and industry (see e.g. \cite{madan2010cell,AndLoadCommag13} and references therein). The implicit assumption in most of these prior works is that a UE may associate with at most one of the available APs/RATs. But given the dense deployment of these networks, the UEs  having simulateneous connectivity to multiple  APs,  may simultaneously associate with multiple RATs and aggregate traffic. As a result, the user throughput and consequently the Quality of Experience (QoE) could be significantly improved. In fact, such architectures are being  standardized in the context of cellular (LTE) and wireless LAN (WLAN) HetNets or LWA\cite{3gpp_r2156737} in LTE Release 13. Furthermore such an architecture is already in place for simultaneous association and traffic splitting across an anchor and booster cell in LTE (or dual connectivity) \cite{Zak13}.  

However, techniques to realize  capacity gains enabled by such architectures are still in nascent stages. Algorithms for traffic aggregation in multi-RAT HetNets have been considered in recent works \cite{GoyVT15,Muk14}. The work in \cite{GoyVT15} investigates the flow allocation for minimizing average delay in co-located multi-RAT setting with no backhaul delay, whereas \cite{Muk14} incorporates backhaul delay in its proposal. Both these works, however, focus on optimal routing from a single UE  perspective and fairness across UEs is not  captured.  The algorithms proposed in  \cite{Sou08,Bethan14} were aimed for  multi-band aggregation and massive MIMO networks respectively and, hence, do not apply directly extend to the multi-RAT aggregation of this paper. 
Traffic (bearer) splitting is an important feature in recent standards such as dual connectivity/LWA, and a low complexity  solution  that works across all deployment scenarios (with and without backhaul delay) and accounts for multi-user fairness is of importance. This paper bridges this gap and proposes a closed form solution which is optimal for a system setting similar to  dual-connectivity and LWA and leverages existing signaling/feedback mechanisms. 
In this paper, we propose and demonstrate a simple yet optimal algorithm for traffic splitting and aggregation in multi-RAT HetNets, where each UE's traffic is split across a macrocell (anchor) and small cell\footnote{Low power APs like LTE femtocells, wireless LAN (WLAN) APs, etc. are generically referred to as small cells in this paper.} (booster). The proposed algorithm  maximizes network wide  proportional fairness through maximizing the sum log throughput across all UEs.  The proposed solution takes into account each UE's spectral efficiency on the macrocell, rate on the small cell,   and  the backhaul delay on small cell. The developed solution is shown to have a tractable  and intuitive ``water-filling" based interpretation, where the water level for a UE is the fraction of resources allocated on macrocell and is inversely proportional to the ratio of small cell  rates and macrocell spectral efficiency.  Using comprehensive LTE-WLAN based simulations, that are complaint with 3GPP evaluation methodology,   the throughput  and capacity gains from the proposed aggregation algorithm are shown to be up to $70\%$ over the baseline RAT association algorithms.  The demonstrated  system  performance gains  provides motivation for the operators to introduce aggregation solutions over their existing small cells/WLAN deployments.

\section{System Model}\label{sec:sysmodel}
\begin{figure}
  \centering
{\includegraphics[width= \columnwidth]{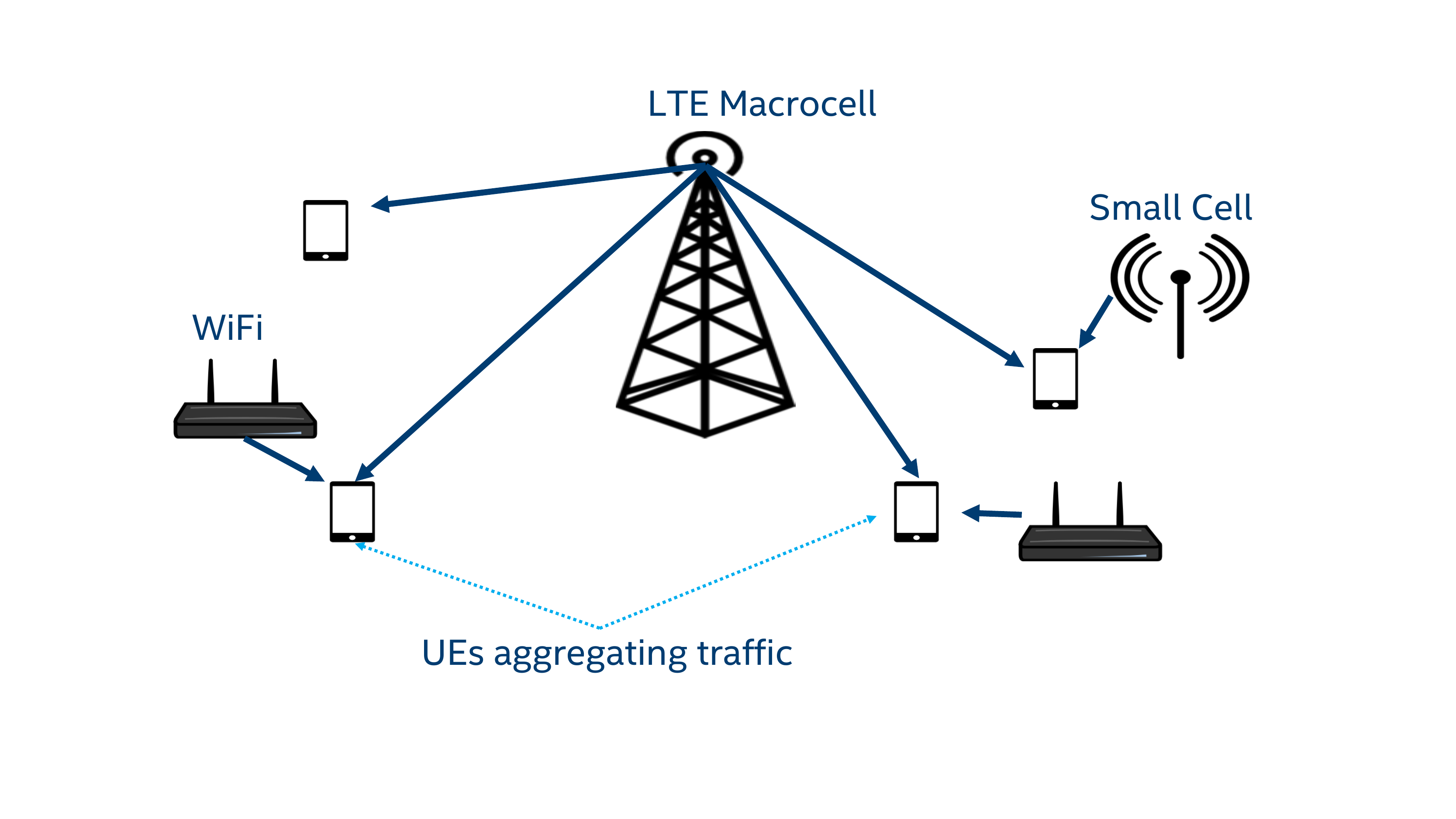}}
\caption{UEs aggregating traffic across LTE macrocell (anchor) and small cells (boosters). }
 \label{fig:sysmodel}
\end{figure}

A HetNet setting  is considered, where multiple small cell APs are deployed within the coverage area of each macrocell.   All the UEs are assumed to have the capability to aggregate traffic over macrocell and small cells. The small cells (e.g. WLAN APs) are assumed to operate on different frequency from that of macrocells. 
 The split of traffic occurs at the macrocell. The instantaneous transmission rate for a UE $k$ from the small cell is  denoted by $\rate_k$, and  is assumed to be known at macrocell from information exchange over the  backhaul\footnote{ Note that such a feedback procedure is already in place for dual-connectivity and LWA scenarios \cite{3gpp_r2156967}.}. The delay on the wired backhaul from macrocell to small cell is denoted by  $\latency$ (typically technology specific), which affects the propagation delay of the traffic sent through small cell  as well as the delay on the reporting of $\rate_k$ to macrocell. The considered scenario is also referred to as the anchor-booster framework \cite{Zak13}, where the LTE macrocell  is the anchor and the small cells are the boosters. An example system model is shown in Fig. \ref{fig:sysmodel}, where UEs aggregate traffic across  macrocell,  small cells, and WLAN APs.

The spectral efficiency $\se_k$ of  user $k$ on the macrocell  is  assumed to be known (with some delay) at the corresponding macrocell using CQI/CSI feedback. The bandwidth of macrocell network is assumed to be denoted by $\res$. For notational brevity, the peak capacity of a UE on macrocell is denoted by $\Pse_k \triangleq \res\se_k$ henceforth. The fraction of resources allocated to  UE $k$ from macrocell is denoted by $\af_k$.

\section{Problem Formulation}
In this paper, we aim to maximize the sum of logarithm of rates over all \textit{active} UEs (i.e. those  with downlink traffic) sharing the resources at macrocell (e.g. all UEs within a sector).  Assuming $K$ UEs in a sector indexed by $k=1\ldots K$, the problem  is to
\begin{subequations}\label{eq:opt}
\begin{align}
\text{maximize } & \sum_{k=1}^K \log(\rate_{\mathrm{eff},k} + \af_k \Pse_k) \label{eq:opt1} \\
\text{subject to } & \sum_{k=1}^K \af_k = 1.  \label{eq:opt2}
\end{align}
\end{subequations}
where $\rate_{\mathrm{eff},k}$ is the effective rate on small cell for $\uth{k}$ UE.  Due to the backhaul delay ($\latency$), the total time required to download a file of size $f$ at UE $k$ is 
$t = l + \frac{f}{r_k}.$
As a result, the effective rate experienced by user $k$ on small cell is  
$\rate_{\mathrm{eff},k} \triangleq \frac{f}{t} = \left(\frac{1}{\rate_k} + \frac{\latency}{f}\right)^{-1}.$
 Note for UEs outside the coverage of any small cell AP, $\rate_k = \rate_{\mathrm{eff},k}  =0$. The file size $f$ may be computed based on the size of the new data that arrives at the common buffer before splitting.

The objective in  (\ref{eq:opt}) is aimed to provide proportional fairness across all UEs associated with the anchor cell (macrocell here)  including those not served by any small cell/booster  subject to the sum resource  constraint at the macrocell. 
 In order to realize the summation of rates across RATs  for each UE, as denoted in (\ref{eq:opt1}), each UE's file is split in the ratio of the corresponding rates at the macrocell and small cell, i.e. the fraction of traffic sent over macrocell for user $k$ is $\frac{\af_k\Pse_k }{\af_k\Pse_k+\rate_{\mathrm{eff},k}}$. 

\section{Optimal Solution and Implementation}
The problem of resource allocation posed in the previous section allows an attractive (simple) and optimal  solution akin to the ``water-filling" solutions in power allocation problems. Using Lagrange multiplier, the problem in (\ref{eq:opt1}) can be expressed as
\begin{align}
\text{maximize } \sum_{k=1}^K \log(\rate_{\mathrm{eff},k} + \af_k \Pse_k)  + \lm\left(\sum_{k=1}^K \af_k - 1\right).
\end{align}
Differentiating with respect to the allocation fraction $\af_k$ and setting to zero, gives 
\begin{equation}
\frac{\Pse_k}{\rate_{\mathrm{eff},k} + \af_k \Pse_k} = -\lm \,\,\,\forall k =1 \ldots K.\end{equation}
Thus, to maximize (\ref{eq:opt1})  subject to (\ref{eq:opt2}),  the condition of 
\begin{align}\label{eq:constfrac}
\frac{\rate_{\mathrm{eff},k}}{\Pse_k} + \af_k = \mathrm{A} \,\,\,\forall k =1 \ldots K,
\end{align}
needs to hold, where $\mathrm{A} = -1/\nu$ is a constant chosen to meet the resource constraint. 
Alternatively, summing over all UEs, the condition becomes
\begin{align}\label{eq:constfrac2}
\frac{1}{K}\left(\sum_{k=1}^K\frac{\rate_{\mathrm{eff},k}}{\Pse_k} + 1 \right) =  \mathrm{A}.
\end{align}

The optimal allocation is obtained by solving the system of linear equations in (\ref{eq:constfrac}) and (\ref{eq:constfrac2}) for   $\eta_k$ and $\mathrm{A}$ using    Algorithm \ref{optalgo}. The algorithm  iteratively eliminates (if needed)  resource fraction to  those UEs  that stand to gain the least from aggregation, i.e. those with highest $\frac{\rate_{\mathrm{eff},k}}{\Pse_k}$. As can be observed from (\ref{eq:constfrac}), the allocated resource fraction for a  UE on macrocell should be inversely proportional to $\frac{\rate_{\mathrm{eff},k}}{\Pse_k}$. Due to the sorting required in Algorithm \ref{optalgo}, the computational complexity scales $O(K\log(K))$, as compared to $O(1)$ scaling of per user allocation algorithms in \cite{Muk14,GoyVT15}. 

\begin{figure}
  \centering
{\includegraphics[width= \columnwidth]{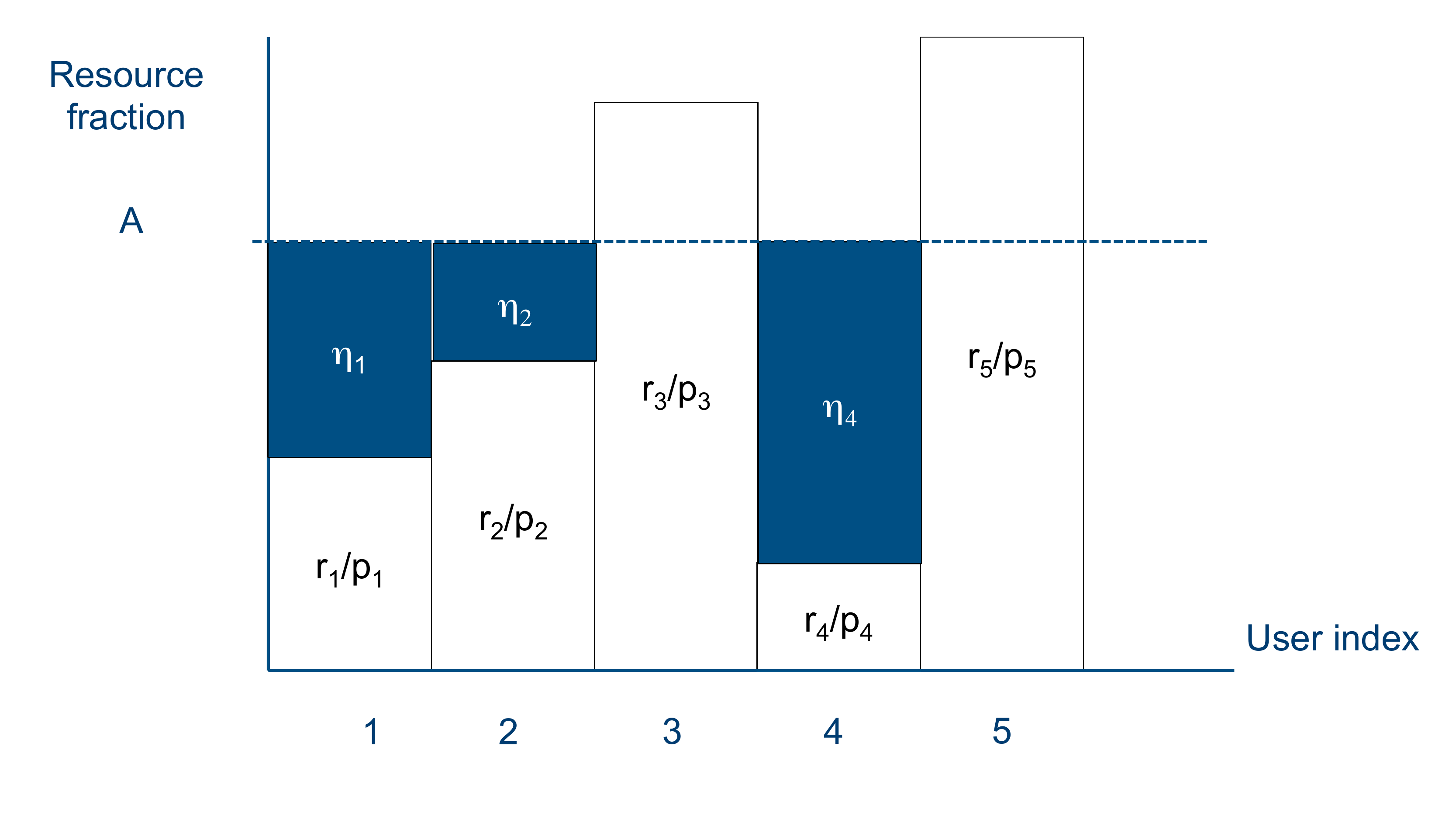}}
\caption{Water filling based interpretation of optimal solution. }
 \label{fig:waterfill}
\end{figure}

The pictorial representation of the algorithm is shown in Fig. \ref{fig:waterfill} for an example setup with five UEs  and their corresponding rate ratios denoted by the height of bars. As seen, the optimal $\af_k$ imitates a ``water-filling" mechanism, where the total amount of water (resources) is distributed across each UEs' cup in a manner such that water poured in any cup is inversely proportional to the height of the cup $\frac{\rate_{\mathrm{eff},k}}{\Pse_k}$. In the shown example, UE $3$ and $5$ are not allocated any resources on macrocell (i.e. $\af_3=\af_5=0$), as the corresponding ratio of small cell throughputs to macrocell capacity is too high to be supported in the current load scenario.

\begin{algorithm}
\caption{Optimal resource fraction algorithm}
\label{optalgo}
\begin{algorithmic}[1]
\Procedure{OPT-ALLOC}{}
\State Sort user indices such that $\rate_{\mathrm{eff},1}/\Pse_1 \leq \rate_{\mathrm{eff},2}/\Pse_2 \leq \ldots \rate_{\mathrm{eff},K}/\Pse_K$
\State  $N= K$, $B= \mathrm{A}$
\While {$\af_N = B - \frac{\rate_{\mathrm{eff},N}}{\Pse_N}  \leq 0$}
\State $N= N-1$
\State $B = \frac{1}{N}\left(\sum_{n=1}^N\frac{\rate_{\mathrm{eff},n}}{\Pse_n} + 1 \right)$
\EndWhile
\State $\af_k = B - \frac{\rate_{\mathrm{eff},k}}{\Pse_k} \,\, \forall k=1\ldots N$; $\af_k = 0 \,\, \forall k=N+1\ldots K$
\State Unsort user indices
\EndProcedure
\end{algorithmic}
\end{algorithm}

\section{Performance Evaluation}
The proposed algorithm is evaluated using a comprehensive system level LTE-WLAN simulator, with evaluation methodology that is 3GPP complaint\cite{3gpp_tr36814}. Specific assumptions and details on WLAN modeling is captured in Table \ref{table:notationtable}. The proposed algorithm is compared with two RAT selection algorithms: \begin{inparaenum} \item WLAN Preferred (WP) and \item LTE Release 12 based radio inter-working solution (Rel12) \cite{TR834}\end{inparaenum}, and a delay equalizing (DE) traffic aggregation algorithm.  In WP, a UE associates with the strongest (in terms of $\SNR$) WLAN  AP given that the corresponding $\SNR$ exceeds a threshold. In the presented results this threshold is assumed to be the minimum $\SNR$ required for  decoding the 1/2 BPSK with $10\%$ PER. In Rel12, a UE associates with the  WLAN AP (using WP algorithm) only when the $\SINR$ from LTE macrocell is below a certain threshold. The optimal value for this threshold is empirically found for the comparisons. Rel12 can, thus, be interpreted as somewhat similar to the biased cell association,   studied in \cite{SinDhiAnd13},  with a large enough bias applied to WLAN APs.   In DE, the traffic for each UE is split in the ratio so as to equalize the packet delay across the two RATs and hence minimize the maximum delay per UE as in \cite{Muk14,GoyVT15}. All these three algorithms, do not account for multi-user fairness across RATs, but they do use ``local" resource allocation algorithms at each RAT based on the corresponding schedulers (as per Table \ref{table:notationtable}).

 The load in the network is varied by varying the number of UEs per sector. The number of $10$, $20$ and $30$ UEs/sector correspond to approximately $20\%$, $40\%$, and $60\%$ network utilization  respectively in our setup.

\begin{figure}
  \centering
{\includegraphics[width= \columnwidth]{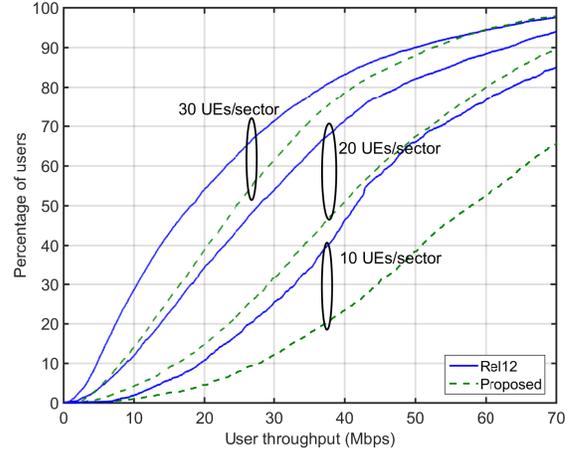}}
\caption{Comparison of user throughput CDF  obtained from  Rel12 and the proposed algorithm with zero delay backhaul for  different load in the network. }
 \label{fig:tputcomp}
\end{figure}

\begin{figure}
  \centering
{\includegraphics[width= \columnwidth]{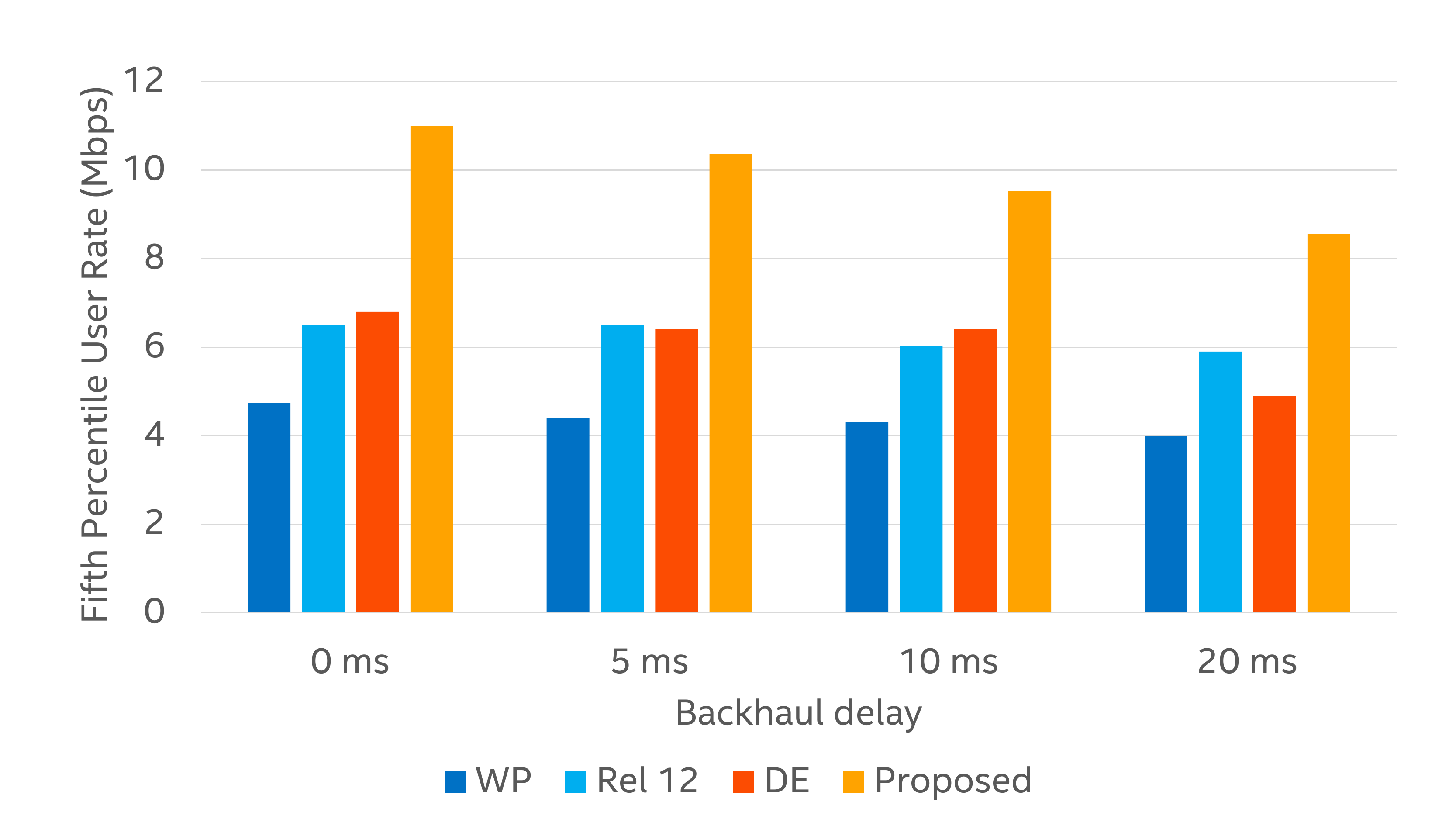}}
\caption{Fifth percentile throughput for different algorithms for varying backhaul delay with $20$ UEs/sector.}
 \label{fig:edgerate}
\end{figure}

\begin{figure}
  \centering
{\includegraphics[width= \columnwidth]{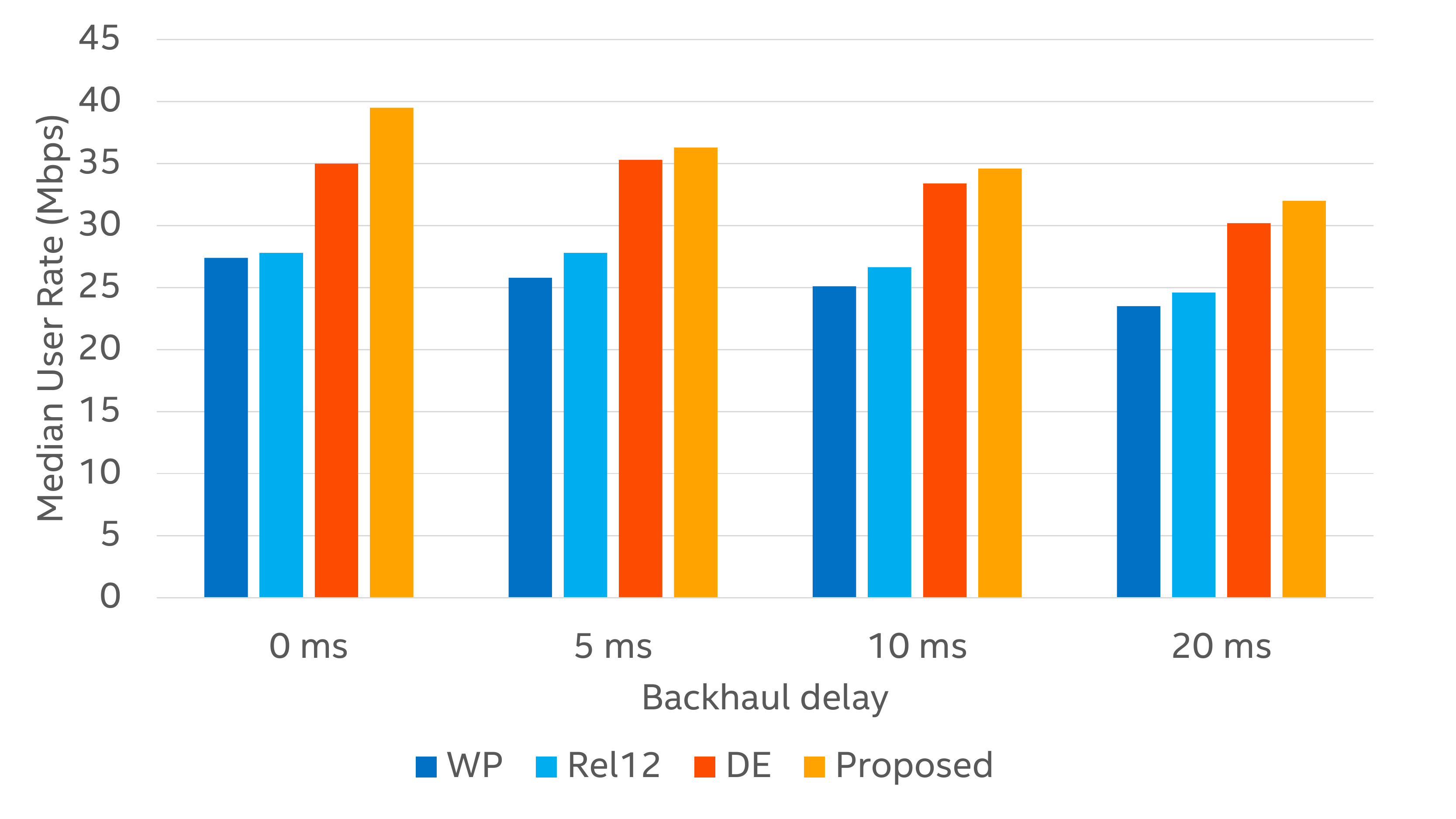}}
\caption{Median throughput for different algorithms for varying backhaul delay with $20$ UEs/sector.}
 \label{fig:medianrate}
\end{figure}

\paragraph{Gains with ideal backhaul}
The distribution of per UE throughput in the network is shown in Fig.~\ref{fig:tputcomp} for the proposed algorithm and Rel12 association algorithm for a zero delay backhaul, i.e., $\latency =0$. As can been seen, for all the shown  load scenarios, the proposed algorithm outperforms Rel12 significantly. In particular,  the proposed algorithm provides the same edge rate ($5-10$ percentile rate) with $30$ UEs/sector as that provided by Rel12 with $20$ UEs/sector, which in turn  implies a $50\%$ gain from a network capacity perspective. 
\paragraph{Gains without ideal backhaul}
  The fifth percentile (or edge)  rate   and median rates obtained from the four algorithms are shown in Fig. \ref{fig:edgerate} and \ref{fig:medianrate} respectively for different backhaul delays. There are three key  observations to be derived from these two figures: \begin{inparaenum} \item as expected both the edge and median rates decrease (across all algorithms) with increasing delay on WLAN backhaul; and \item the proposed aggregation  provides a gain of about $60$-$70\%$ in the edge rates and $30$-$40\%$ in median rates (across all delays) over Rel12 based selection algorithm; and \item the proposed algorithm performs much better than DE in edge rates than that in median rates  because of  the multi-user proportional  fairness aspect incorporated in the developed algorithm\end{inparaenum}. Thus, as substantiated by these results, the proposed algorithm is well suited for LTE-WLAN  aggregation with low as well as large backhaul delay.

\begin{table}
	\centering
\caption{Simulation Assumptions}
	\label{table:notationtable}
  \begin{tabulary}{\linewidth}{|L|L| }
    \hline
    \textbf{Notation} & \textbf{Description} \\\hline
    Topology &	3 sectors per macrocell, 5 small cells/sector, 7 cell wrap-around \\\hline
UE dropping; UE density	& Clustered; $10$, $20$, $30$ UE per sector\\\hline
Traffic & Non full buffer with exponentially distributed inter-arrival  time with mean $1$s, and fixed file size of $0.5$ MB/file across all UEs (3GPP FTP traffic model 3) \\\hline
LTE Carrier Frequency; channel; UE speed  & $2$ GHz;	[IMT] UMa Macro, UMi Pico; UE speed= $3$ km/hr  \\\hline
LTE mode &	Downlink FDD; $20$ MHz for DL   \\\hline
No. antennas (macro, UE)	 & ($2$,$ 2$) \\\hline
Antenna configuration	& macrocell: co-polarized, UE: co-polarized    \\\hline
Max rank per UE	& $2$ (SU-MIMO)  \\\hline
UE channel estimation; Feedback/control channel errors	& Ideal; no Error \\\hline
PHY Abstraction &	Mutual information  \\\hline
Scheduler; LTE scheduling granularity		& Proportional fair scheduler for LTE and round-robin for WLAN;  $5$ PRBs \\\hline
Receiver type	& Interference unaware MMSE  \\\hline
CQI and PMI feedback granularity  in frequency; Feedback periodicity&  $5$ PRBs;  $10$ms   \\\hline
PMI feedback	& 3GPP Rel-10 LTE codebook (per sub-band)  \\\hline
Outer loop for target FER control &	$10\%$ PER for 1st transmission  \\\hline
Link adaptation	& MCSs based on LTE transport format  \\\hline
HARQ scheme &	CC  \\\hline
    \multicolumn{2}{|c|}{WLAN} \\\hline  
WiFi deployment & IEEE 802.11n based APs, uniform deployment within sector  \\\hline
WiFi frequency; channelization 	& $2.4$ GHz band, $3$ frequency bands, $20$ MHz channels; least power based channel selection   \\\hline
AP transmit power; TX-OP limit &	$20$ dBm; 	$1$ms  \\\hline
PHY Abstraction; MPDU size  &	RBIR; $1500$ bytes   \\\hline
\end{tabulary}
\end{table}

\section{Conclusion}

To the authors' best knowledge, this is the first work to propose and demonstrate an algorithm for aggregating traffic in LTE-WLAN HetNets with non-ideal backhaul while accounting for multi-user fairness.  Moreover,  the paper is also one of the first  to benchmark performance gains achievable with the  recently developed 3GPP LWA framework  compared to current 3GPP-WLAN interworking solutions for practical deployment scenarios of interest to network operators.  
The developed framework is also applicable to  RATs employing  millimeter-wave based frequencies as well. Future work could extend stochastic geometry based analysis, e.g. \cite{SinDhiAnd13}, could also be extended to investigate coverage and capacity in HetNets with  traffic splitting and aggregation.

%
\end{document}